\documentclass[conference]{IEEEtran}
\ifCLASSINFOpdf
  \usepackage[pdftex]{graphicx}
  \usepackage{url}
\else
  \usepackage[dvips]{graphicx}
  \graphicspath{{../figs/}}
  \DeclareGraphicsExtensions{.eps}
\begin{document}
%
\title{Modified Orthogonal Matching Pursuit Algorithm for Cognitive Radio Wideband Spectrum Sensing}

\author{\IEEEauthorblockN{Peng Zhang, Robert C. Qiu}
\IEEEauthorblockA{Department of Electrical and Computer Engineering\\
Cookeville, TN 38505\\
Tennessee Technological University\\
Email: $\{$pzhang21,rqiu$\}$@tntech.edu}}

\maketitle

\begin{abstract}
Sampling rate is the bottleneck for spectrum sensing over multi-GHz bandwidth. Recent progress in compressed sensing (CS) initialized several sub-Nyquist rate approaches to overcome the problem. However, efforts to design CS reconstruction algorithms for wideband spectrum sensing are very limited. It is possible to further reduce the sampling rate requirement and improve reconstruction performance via algorithms considering prior knowledge of cognitive radio spectrum usages. In this paper, we group the usages of cognitive radio spectrum into three categories and propose a modified orthogonal matching pursuit (OMP) algorithm for wideband spectrum sensing. Simulation results show that this modified OMP algorithm outperforms two modified basis pursuit de-noising (BPDN) algorithms in terms of reconstruction performance and computation time. 
\end{abstract}

\section{Introduction}
\label{introduction}
\begin{figure}
	\centering
		\includegraphics[width=0.40\textwidth]{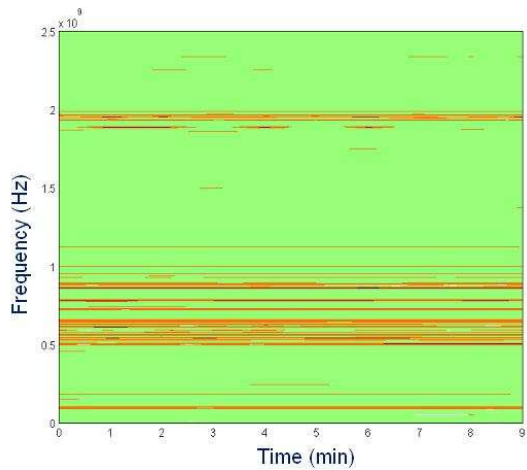}
	\caption{Spectrum utilization measured at Berkeley wireless research center.}
	\label{fig:spectrum_usage}
\end{figure}

Although the spectrum is almost fully allocated to various radio frequency (RF) technologies, it is not fully utilized. The spectrum measurement result in Fig. \ref{fig:spectrum_usage} \cite{bwrcspectrum} shows that most of time, the utilization of the spectrum over $2.5$ GHz bandwidth can be as low as $10\%$. This utilization is even lower in rural areas. Cognitive radio is the idea to increase the utilization. Spectrum sensing, the operation to find the unused spectrum holes in the whole spectrum, is essential to cognitive radio. Spectrum sensing for bandwidth over multi-GHz, however, posed a challenge to today's sampling technology. According to Nyquist sampling theorem, the sampling rate should be at least twice of the signal bandwidth in order to reconstruct the original signal. Current analog-to-digital converter (ADC) to reach multi-GHz sampling rate is hardly affordable. 

To reduce the sampling rate requirement, compressed sensing (CS) \cite{donoho2006cs,candes2006cs} seems to be a natural approach. According to encouraging results from CS theory, the lowest sampling rate for perfect reconstruction is related to the sparsity, a.k.a. number of non-zeros, of the original signal. The sampling rate can therefore be reduced greatly compared with Nyquist rate. The CS theory has been successfully applied in many aspects of communications to reduce sampling rate \cite{Zhang2009,ArceSadler2007CDUWB,ParedesArce2007UWBCS}. In cognitive radio, the spectrum is sparse in the frequency domain and CS has been applied to overcome the sampling rate problem.  Several sampling structures have been proposed \cite{tian2007compressed,polo2009cwbss,mishali2009theory}. These sampling structures have made CS based spectrum sensing possible, even in hardware prototypes \cite{AIC_2008,xampling_2009}. In this paper, however, we focus on reconstruction algorithms.

Standard CS algorithms have been used to reconstruct the original spectrum, such as basis pursuit (BP) \cite{chen2001atomic}, orthogonal matching pursuit (OMP) \cite{tropp2007srr}, etc. We propose to modify the standard OMP algorithm by using prior knowledge of cognitive radio spectrum usages to improve the reconstruction performance, thus further reducing the sampling rate requirement. There are some cooperative wideband spectrum sensing algorithms modified upon standard CS algorithms \cite{zhang2010stbcs,ying2009dcw}, using shared knowledge of secondary users, under the assumption that there is channel to share information. In this paper, we do not assume secondary users to have channel for shared knowledge, and only prior knowledge for each individual secondary user is available. Very limited efforts have been made in modifying CS algorithms with prior knowledge for wideband spectrum sensing. An obvious prior knowledge of cognitive radio spectrum usages is fixed spectrum boundaries set by FCC for primary users. Such structural information can be used in CS algorithms as prior knowledge to improve the reconstruction performance. `Mixed $l_2 / l_1$ norm denoising operator' (MNDO) \cite{liu2010compressive} is an algorithm that modifies the BP problem formulation using such prior knowledge. Another BP based algorithm developed for magnetic resonance imaging (MRI), called modified BPDN \cite{modifiedbpdn2010}, has similar idea, and is introduced by us in wideband spectrum sensing for comparison purpose. 

In this paper, we re-examine the prior knowledge of cognitive radio spectrum usages and propose a modified OMP algorithm. Similar to \cite{haykin2005cognitive}, the spectrum usages in a cognitive radio network can be grouped into three categories as listed in Table \ref{tab:categories}.
\begin{table}
	\centering
	\caption{Three Categories of Cognitive Radio Spectrum Usage}
	\label{tab:categories}
		\begin{tabular}{|p{8 cm}|}
	1. Spectrum bands with fixed boundaries which are always accessed by primary users. Example: Local radio station signals, local TV signals, etc.\\
  2. Spectrum bands with fixed boundaries which are accessed by primary users not very often. Example: TV spectrum available for white space device access \cite{FCC2008WSD}.\\
  3. Spectrum bands with fixed boundaries which are partially and randomly accessed. Example: Cellphone signals.
		\end{tabular}
\end{table}
The modified OMP algorithm considers all three categories. We will show by simulation that the modified OMP algorithm has much better reconstruction performance in mean square error (MSE) than MNDO and modified BPDN, given the same amount of spectrum utilization. Moreover, our algorithm takes much less computation time compared with MNDO and modified BPDN. Such improvements are benefited from proper use of prior knowledge and the greedy nature of OMP algorithm. 

This paper is organized as follows. Section \ref{systemmodel} states the wideband spectrum sensing model under the CS framework. Section \ref{MOMP} introduces the MNDO, modified BPDN and our proposed modified OMP algorithm. Simulation results are shown in Section \ref{simulation} and conclusions are made in Section \ref{conclusion}.

\section{Wideband Spectrum Sensing Model via Compressed Sensing}
\label{systemmodel}

\subsection{Discrete Spectrum Model}
\label{spectrummodel}
Suppose that the spectrum available for cognitive radio network has total bandwidth of $B$ Hz. It is convenient to describe the spectrum in discrete domain using an $N \times 1$ frequency domain sample vector ${\bf f}$:
\begin{equation}
\label{freqvector}
{\bf f} = \left[ {\begin{array}{*{20}c}
   {f_1 } & {f_2 } & {...} & {f_N }  \\
\end{array}} \right] ^T
\end{equation}
where T is transpose, $\{ f_i \}$ are signal values in frequency domain uniformly sampled over $B$ with spacing $B / N$, and indices $\{i\}$ are related to frequency locations. In noiseless case, if $\left| {f_i } \right|^2  \ne 0$, then the spectrum is occupied at the $i$-th frequency location. Otherwise, $i$ represents the unused frequency location that is accessible to secondary users. The frequency sample vector ${\bf f}$ can be projected onto Nyquist rate time domain sample vector ${\bf r}$ via an inverse $N \times N$ Fourier projection matrix ${\bf F}^{-1}$:
\begin{equation}
	{\bf r} = {\bf F}^{-1}{\bf f}
\label{freqmodel}
\end{equation}

According to the three categories mentioned in Table \ref{tab:categories}, fixed spectrum boundaries are known in prior. Indices $\{b_i\}$ denote frequency boundaries and they can separate the whole spectrum into $K$ consecutive subsections: 

\begin{equation}
\label{subsections}
\begin{array}{l}
 u_1  = \left\{ {1,2,...,b_1 } \right\} \\ 
 u_2  = \left\{ {b_1  + 1,b_1  + 2,...,b_2 } \right\} \\ 
 ... \\ 
 u_K  = \left\{ {b_{K - 1}  + 1,b_{K - 1}  + 2,...,N} \right\} \\ 
 \end{array}
\end{equation}

Now, we define three category sets $\{S_n\}$ according to the following conditions:

\begin{equation}
\label{subsets}
\begin{array}{l}
 S_n  = \left\{ {u_i \left| {i \in {\rm Category }\;n} \right.} \right\},n = 1,2,3 \\ 
 {\bf \Omega } = \bigcup\limits_n {S_n } \; \; \left( S_i  \cap S_j  = \emptyset ,{\rm  for }\;i \ne j \right)\\ 
 \end{array}
\end{equation}
where ${\bf \Omega}$ is the universal set.

According to the measurement results of spectrum utilization, we assume the following relation is valid

\[
| {\bf f} | /N \leq 10\%
\]
where $| {\bf f}  |$ is the number of non-zeros in ${\bf f}$.

\subsection{Compressed Sensing Based Spectrum Sensing Approach}

In the framework of CS \cite{candes2006nos,donoho2006cs}, consider the sampling of an $N \times 1$ signal vector ${\bf x} = {\bf \Psi} {\bf s}$, where ${\bf \Psi}$ is an $N \times N$ dictionary matrix and ${\bf s}$ is an $N \times 1$ vector with $L << N$ non-zero entries $s_i$. It has been shown that ${\bf x}$ can be recovered with $M$ samples by projecting ${\bf x}$ via an $M \times N$ measurement matrix $\bf \Phi$, where $L < M < N$, and ${\bf \Phi}$ is incoherent with ${\bf \Psi}$. The $M \times 1$ measurement vector $\bf y$ and the projection process can be represented as

\begin{equation}
\label{CSprojection}
	{\bf y = \Phi x = \Phi \Psi s}	
\end{equation}

$\bf s$ can be reconstructed by BP \cite{chen2001atomic}, which solves the following $l_1$ norm convex optimization problem:
\begin{equation}
\label{l1}
\begin{array}{l}
{\bf \hat s} = \arg \mathop {\min }\limits_{\bf s} \left\| {\bf s} \right\|_1   \\ 
 {\rm s}{\rm .t}{\rm .}\;\;{\bf \Phi \Psi s = y} \\ 
 \end{array}
\end{equation}
where $l_p$ norm of $\bf s$ for $p \geq 1$ is defined as:
\[
\left\| {\bf s} \right\|_p  = \left( {\sum {|{\bf s}_i| ^p } } \right)^{\frac{1}{p}} 
\]

The signal can also be reconstructed by greedy algorithm like OMP \cite{tropp2007srr}, which will be described in detail.

If we compare the system model in Equation (\ref{freqmodel}) and Equation (\ref{CSprojection}), received time domain sample vector $\bf r$ can be considered as the signal vector $\bf x$, inverse Fourier projection matrix ${\bf F}^{-1}$ can be considered as the dictionary matrix $\bf \Psi$ and frequency domain sample vector $\bf f$ can be considered as the sparse signal vector $\bf s$. If a proper implementable measurement matrix $\bf \Phi$ is designed to be incoherent with ${\bf F}^{-1}$, then the spectrum sensing problem can be modeled as the CS reconstruction problem, and sub-Nyquist rate sampling rate reconstruction can be possible via CS algorithms. As mentioned in Section \ref{introduction}, several efforts have been made for implementable measurement approaches in real world. In this paper, we focus on the reconstruction algorithm that can be applied to any of the sub-Nyquist rate measurement approaches. To emphasize the performance of reconstruction algorithms, we simply use an $M \times N$ Gaussian random matrix $\bf \Phi$ to represent the measurement process together with inverse Fourier projection matrix ${\bf F}^{-1}$. The use of Gaussian random matrix can guarantee reconstruction performance \cite{candes2006rup}. The simplified spectrum sensing model is:

\begin{equation}
	\label{simplifiedproblem}
	{\bf y = \Phi f}
\end{equation}

As a result, $\bf f$ can be reconstructed by solving the problem
\begin{equation}
\label{simplifiedl1}
\begin{array}{l}
{\bf \hat f} = \arg \mathop {\min }\limits_{\bf f} {\left\| {\bf f} \right\|_1 }\\ 
 {\rm s}{\rm .t}{\rm .}\;\;{\bf \Phi } {\bf f = y} \\ 
 \end{array}
\end{equation}

For a more general model with additive white Gaussian noise (AWGN), we have
\begin{equation}
	\label{awgnmodel}
	{\bf y = \Phi f + w}
\end{equation}
where $\bf w$ is $M \times 1$ noise vector with normal distribution. The CS problem in (\ref{simplifiedl1}) can be modeled using BPDN \cite{chen2001atomic}:
\begin{equation}
\label{awgnl1model}
{\bf \hat f} = \arg \mathop {\min }\limits_{\bf f} \frac{1}{2}\left\| {{\bf y} - {\bf \Phi }{\bf f}} \right\|_2 + \gamma \left\| {\bf f} \right\|_1 
\end{equation}
where $\gamma$ is determined by noise level. (\ref{simplifiedl1}) is a special case of (\ref{awgnl1model}) by setting $\gamma$ to $0$.

In the following sections, we will compare different reconstruction algorithm performance using the simplified model in Equation (\ref{awgnmodel}). 

\section{Modified Orthogonal Matching Pursuit}
\label{MOMP}

\subsection{BP Based Reconstruction Algorithms Considering Prior Knowledge of Cognitive Radio Spectrum}
\label{crmodel}

In order to consider the prior knowledge described in Table \ref{tab:categories}, we need to re-formulate the $l_1$ norm problem in Equation (\ref{awgnl1model}). 

MNDO \cite{liu2010compressive} considered spectrum allocation boundaries as prior knowledge. The frequency domain sample vector $\bf f$ is divided into `blocks' according to boundary information $\{ b_i \}$:

\begin{equation}
	\label{blockvector}
{\bf f} = \left[ \begin{array}{l}
 \underbrace {\begin{array}{*{20}c}
   {f_1, } & {...,} & {f_{b_1 } ,}  \\
\end{array}}_{{\bf f}_1 }\underbrace {\begin{array}{*{20}c}
   {f_{b_1  + 1}, } & {...,} & {f_{b_2 }, }  \\
\end{array}}_{{\bf f}_2 } \\ 
 ...\underbrace {\begin{array}{*{20}c}
   {f_{b_{K - 1}  + 1}, } & {...,} & {f_{b_K }, }  \\
\end{array}}_{{\bf f}_K } \\ 
 \end{array} \right]^T 
\end{equation}

MNDO re-formulates BP to a new convex optimization problem with boundary information: 

\begin{equation}
\label{blockmodel}
\begin{array}{l}
{\bf \hat f} = \arg \mathop {\min }\limits_{\bf f} \left( {\left\| {{\bf f}_1 } \right\|_2  + \left\| {{\bf f}_2 } \right\|_2  +  \cdots  + \left\| {{\bf f}_K } \right\|_2 } \right) \\ 
 {\rm s}{\rm .t}{\rm .}\;\;\left\| {{\bf y - \Phi} {\bf f}} \right\|_2  \le \eta  \\ 
 \end{array}
\end{equation}
where $\eta$ is determined according to the noise variance. In this problem formulation, all values in the same block are correlated. They are prone to be zeros or non-zeros simultaneously. All blocks are treated evenly, without considering the categories in Table \ref{tab:categories}. As will be shown in simulation in Section \ref{simulation}, noisy signal will be reconstructed over multiple blocks even in noise-free scenario.

Another reconstruction algorithm `modified BPDN' has similar idea, and we propose it for cognitive radio spectrum sensing for comparison purpose. It groups the indices into two sets: 

\begin{itemize}
	\item `Dense' index set $T = \{i\}$ where non-zeros are likely to exist.
	\item `Sparse' index set $T^c$ where non-zeros are not likely to exist and $T^c$ denotes the complement set of $T$.
\end{itemize}

The modified BPDN algorithm models a convex optimization problem using the above prior knowledge is as follows.

\begin{equation}
	\label{modcs}
{\bf \hat f} = \arg \mathop {\min }\limits_{\bf f} \frac{1}{2}\left\| {{\bf y - \Phi } {\bf f}} \right\|_2^2  + \gamma \left\| {{\bf f}_{T^c } } \right\|_1 
\end{equation}
where $\gamma$ is determined by noise variance, ${\bf f}_{T^c}$ denotes a sub-vector containing the elements of $\bf f$ with indices in $T^c$. In the solution of this convex problem, index sets in $T$ are prone to have non-zero values, and index sets in $T^c$ are prone to have zeros values.

We can use modified BPDN for wideband spectrum sensing. From Table \ref{tab:categories}, indices in $S_1$ are prone to have non-zeros while indices in $S_2$ and $S_3$ are prone to have non-zeros. Therefore, it is straightforward to put $S_1$ into dense index set $T$, while $S_2$ and $S_3$ into sparse index set $T^c$. However, if $S_2$ and/or $S_3$ have jointed subsets, some boundary indices will be invalid, resulting in invalid prior knowledge. Here we give a simple example of such invalid prior knowledge. Assume we have ${u_n} = {\rm{ }}\left\{ {{b_{n - 1}} + 1,{b_{n - 1}} + 2...,{b_n}} \right\} \subset {S_2}$ or $S_3$, ${u_m} = {\rm{ }}\left\{ {{b_{m - 1}} + 1,{b_{m - 1}} + 2...,{b_m}} \right\} \subset {S_2}$ or $S_3$, with four boundary indices $b_{n-1}$, $b_{n}$, $b_{m-1}$ and $b_{m}$. If $b_n = b_{m-1}$, then there should be three valid boundary indices $ b_{n-1}$, $b_{n}$ and $b_{n+1}$. However, in the modified BPDN model, $u_n  \cup u_m  = \left\{ {b_{n - 1}  + 1,...,b_n ,b_n  + 1,...,b_{n + 1} } \right\} \subset T^c $, and boundary index $b_n$ becomes invalid. As a result, the modified BPDN will reconstruct signal without considering the boundary index $b_n$ separating $u_n$ and $u_m$.

\subsection{Modifying the OMP Algorithm}
We have introduced two BP based reconstruction algorithms considering prior knowledge and their limitations. To overcome these, we propose a modified OMP algorithm, which will be able to deal with the three categories separately. 

OMP is a greedy algorithm. It reconstructs original signal by iterative search for non-zero indices and performs least-squares estimation of the values on the non-zero indices. The original OMP algorithm \cite{tropp2007srr} is as follows:

\begin{table}
	\centering
	\caption{The Original OMP Algorithm.}
	\label{tab:OMP}

		\begin{tabular}{|p{8 cm}|}
\textsc{Input}:

\begin{itemize}
	\item An $M \times N$ matrix ${\bf \Theta  = \Phi }$
	\item An $M \times 1$ sample vector $\bf y$
	\item Maximum number of iterations $m$
	\item Error tolerance $\eta$
\end{itemize}
\textsc{Output}:

\begin{itemize}
	\item An estimate $N \times 1$ vector $\hat{\bf f}$ for the ideal signal
	\item An index set ${\bf \Lambda} _t$ containing $t$ elements from $\{1,...,N\}$
	\item An $M \times 1$ residual vector ${\bf res}_t$
\end{itemize}
\textsc{Procedure}:

\begin{enumerate}
	\item Initialize the residual ${\bf res}_0 = {\bf y}$, the index set ${\bf \Lambda} _0 = \bf \emptyset$ and the iteration counter $t = 1$. 
	\item Find the index $\lambda_t$ that satisfies the following equation
\begin{equation}
\label{lambdamax}
\lambda _t  = \arg \mathop {\max }\limits_{j = 1,...,N} \left| {\left\langle {{\bf res}_{t - 1} ,{\bf \theta }_j } \right\rangle } \right|
\end{equation}
where ${\bf \theta}_j$ denotes the $j$-th column vector of ${\bf \Theta}$, $\left\langle {{\bf a,b}} \right\rangle$ denotes the inner product of two vectors $\bf a$ and $\bf b$. If the maximum occurs for multiple indices, break the tie deterministically.
	\item Augment the index set ${\bf \Lambda}_t = {\bf \Lambda}_{t-1} \bigcup \{\lambda_t\}$ and the matrix of chosen atoms ${\bf \Theta}_t = {\bf \Theta} _{{\bf \Lambda}_t}$, where ${\bf \Theta} _{{\bf \Lambda}_t}$ is a sub-matrix of $\bf \Theta$ containing columns with indices in ${\bf \Lambda}_t$.
	\item Solve the least-squares problem to obtain a new signal estimate:
\begin{equation}
\label{leastsquare}
{\bf x}_t  = \arg \mathop {\min }\limits_{\bf x} \left\| {{\bf \Theta }_t {\bf x} - {\bf y}} \right\|_2 
\end{equation}
	\item Calculate the new residual:
\begin{equation}
\label{getresidual}
{\bf res}_t  = {\bf y} - {\bf \Theta }_t {\bf x}_t 
\end{equation}
	\item If $t < m$ or $\left\| {{\bf res}_t } \right\|_2  > \eta $, increment $t$, and return to Step 2.
	\item ${\bf x} _t$ is the estimated signal ${{\bf \hat f}}$, with non-zero indices at components listed in ${\bf \Lambda} _t$. 
	
\end{enumerate}

		\end{tabular}
\end{table}

The iteration operation gives us the freedom to consider three categories separately. Since there must be non-zeros in $S_1$, we can initialize the original ${\bf \Lambda}_0$ with $S_1$. This modification benefits the reconstruction accuracy because occupied indices will always be counted. Then, during each iteration, if index $\lambda _t$ is found and if $\lambda_t \in u_i \subset S_2 $ as well, all elements in $u_i$ will be added to ${\bf \Lambda} _t$. This modification benefits the reconstruction accuracy because only selected $\{ u_i \}$ are counted. If index $\lambda _t \in u_i \subset S_3$, only $\lambda _t$ will be added, similar to the original OMP process. As a result, all three categories are treated separately. The modified OMP algorithm is summarized as:

\begin{table}
	\centering
	\caption{The Modified OMP Algorithm.}
	\label{tab:ModOMP}
		\begin{tabular}{|p{8 cm}|}
\textsc{Input}:

\begin{itemize}
	\item An $M \times N$ matrix ${\bf \Theta  = \Phi}$
	\item An $M \times 1$ sample vector $\bf y$
	\item Boundary information $\{ b_1, b_2, ..., b_K \}$, $\{ u_1, u_2, ..., u_K \}$ and $\{ S_1, S_2, S_3 \}$
	\item Maximum number of iterations $m$
	\item Error tolerance $\eta$
\end{itemize}
\textsc{Output}:

\begin{itemize}
	\item An estimate $N \times 1$ vector $\hat{\bf f}$ for the ideal signal
	\item An index set ${\bf \Lambda} _t$ containing $t$ elements from $\{1,...,N\}$
	\item An $M \times 1$ residual vector ${\bf res}_t$
\end{itemize}
\textsc{Procedure}:

\begin{enumerate}
	\item Initialize the residual ${\bf res}_0 = {\bf y}$, the index set ${\bf \Lambda} _0 = S_1$, the matrix of chosen atoms ${\bf \Theta}_1 = {\bf \Theta}_{S_1}$ and the iteration counter $t = 1$.
	\item Solve the least-squares problem in Equation (\ref{leastsquare}) to obtain a new signal estimate.
	\item Calculate the new residual using Equation (\ref{getresidual}).
	\item Increment $t$.
	\item Find the index $\lambda_t$ that satisfies Equation (\ref{lambdamax}).
	\item Augment the index set ${\bf \Lambda}_t = {\bf \Lambda}_{t-1} \bigcup \{\lambda_t\}$. If $\lambda _t  \in u_i  \subset S_2  $, let ${\bf \Lambda}_t = {\bf \Lambda}_t \bigcup u_i$. 
	\item Set the matrix of chosen atoms ${\bf \Theta}_t = {\bf \Theta}_{{\bf \Lambda}_t}$.
	\item Solve the least-squares problem in Equation (\ref{leastsquare}) to obtain a new signal estimate.
	\item Calculate the new residual using Equation (\ref{getresidual}).
	\item Return to Step 4 if $t < m$ or $\left\| {{\bf res}_t } \right\|_2  > \eta $.
	\item ${\bf x} _t$ is the estimated signal ${{\bf \hat f}}$, with non-zero indices at components listed in ${\bf \Lambda} _t$. 

\end{enumerate}

		\end{tabular}
\end{table}

\section{Simulation Results}
\label{simulation}

In the simulation, we use a spectrum from $1$ MHz to $1000$ MHz with approximately $10\%$ utilization to test the performances of the reconstruction algorithms. There are $1000$ discrete frequency locations with $1$ MHz resolution. Number of original siganl samples is $N = 1000$. Assume boundary information and category information as prior knowledge. Using Equation (\ref{subsections}) and (\ref{subsets}), we define three categories $S_1, S_2, S_3$ for demonstration purpose:

\[
\begin{array}{l}
 {\bf \Omega } = \left\{ {1,2,...,1000} \right\} = S_1  \cup S_2  \cup S_3  \\ 
 S_1  = \left\{ {k_1 ,k_2 ,...,k_5 } \right\} \\ 
 \end{array}
\]
where 
\[
\begin{array}{l}
 k_1  = \left\{ {11,12,...,20} \right\} \\ 
 k_2  = \left\{ {265,266,...,274} \right\}\\ 
 k_3  = \left\{ {276,277,...,285} \right\}\\ 
 k_4  = \left\{ {601,602,...,610} \right\}\\ 
 k_5  = \left\{ {701,702,...,710} \right\} \\ 

 \end{array}
\]

\[
S_2  = \left\{ {l_1 ,l_2 ,...,l_{15},l_{18},l_{19},...,l_{40} } \right\}
\]
where 

\[
\begin{array}{l}
 l_i  = \left\{ \begin{array}{l}
 100 + 11\left( {i - 1} \right), 100 + 11\left( {i - 1} \right) + 1, ...,100 + 11i - 1 \\ 
 \end{array} \right\}, \\ 
 i = 1,2,...,40 \\ 
 S_3  = {\bf \Omega }/\left\{ {S_1  \cup S_2 } \right\} \\ 
 \end{array}
\]

Note that this setup is just one example. It can be changed according to specific frequency ranges and frequency allocations.

\begin{figure}[htbp]
	\centering
		\includegraphics[width=0.50\textwidth]{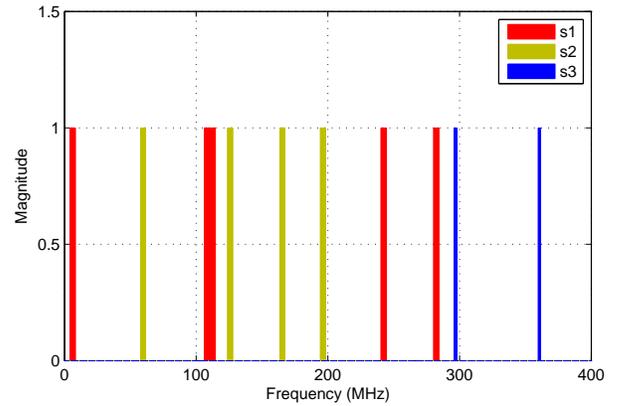}
	\caption{One realization of the spectrum utilization.}
	\label{fig:alpha}
\end{figure}

When building up the spectrum utilization, all frequency locations in $S_1$ are occupied, $10\%$ frequency locations in $S_2$ are randomly occupied, and $2\%$ frequency locations in $S_3$ are randomly occupied. The total number of non-zeros is controled around $100$. Fig. \ref{fig:alpha} illustrates one realization of the spectrum. We set the non-zero values to $1$ for better illustration. The reconstruction algorithms can deal with arbitrary values.

\begin{figure}[htbp]
	\centering
		\includegraphics[width=0.50\textwidth]{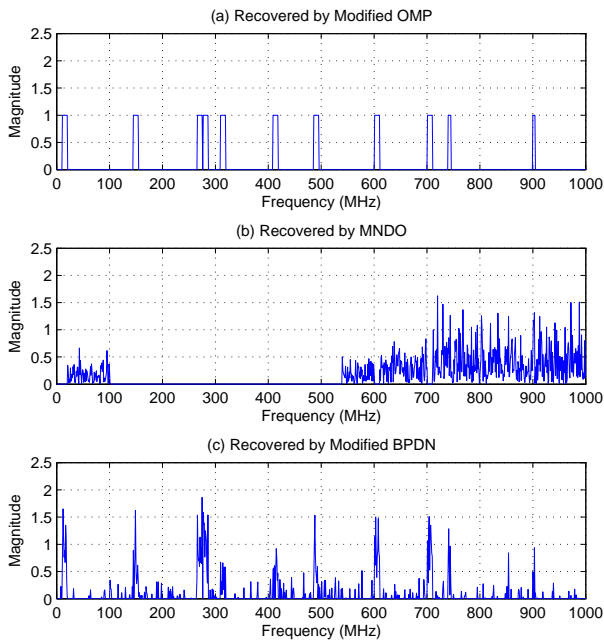}
	\caption{(a) Recovered spectrum by modified OMP. (b) Recovered spectrum by MNDO. (c)  Recovered spectrum by modified BPDN}	
	\label{fig:NoiselessCase}
\end{figure}

We first demonstrate the sub-Nyquist rate reconstruction performance using Fig. \ref{fig:alpha}. Let number of measurements $M = 25\% N = 250$ and let $\bf \Phi$ be $250 \times 1000$ Gaussian random matrix. Fig. \ref{fig:NoiselessCase} shows the results from modified OMP, MNDO and modified BPDN, under the noiseless case. We use normalized MSE as a metric to evaluate their performances:
\[
{\rm{MSE}} = \frac{{{{\left\| {{\bf{\hat f}} - {\bf{f}}} \right\|}_2}}}{{{{\left\| {\bf{f}} \right\|}_2}}}
\]
MNDO has MSE $\approx 1.8908$. From Fig. \ref{fig:NoiselessCase}, we can see that noisy spikes are generated over multiple frequency blocks within $50 - 200$ MHz and $1100 - 2000$ MHz. Modified BPDN has MSE $\approx 0.3331$. Since boundary indices of jointed subsets become invalid, non-zero spikes can be found throughout the spectrum. Modified OMP, however, is able to reconstruct the original signal almost perfectly with MSE $\approx 10^{-30}$.

Now we use Monte Carlo simulation to compare the reconstruction performance for noiseless signal with $M = 250$ and $M = 500$. $500$ different spectrum realizations are generated for the simulations. The averaged MSE performances for modified OMP, MNDO and Modified BPDN are in Table \ref{tab:noiselesscase}. It can be seen that when $M = 250$, only modified OMP can provide almost perfect reconstruction with MSE $\approx 1^{30}$. When $M = 500$, modified BPDN can reach almost perfect reconstruction with MSE $\approx 10^{-20}$, while MNDO still has a few errors with MSE $\approx 0.081$.

\begin{table}[htbp]
	\centering
	\caption{Averaged MSE for Noiseless Spectrum Realizations}
	\label{tab:noiselesscase}
		\begin{tabular}{|c||c|c|}
		\hline
		 & M = $250$ & M = $500$\\ \hline
		MNDO & $1.9975$ &$0.0810 $\\ \hline
		Modified BPDN & $0.1438 $& $7.920 \times 10^{-20}$ \\ \hline
		Modified OMP & $10 ^{-30}$ & $7 \times 10^{-31} $\\ \hline
	\end{tabular}
\end{table}

\begin{figure}[htbp]
	\centering
		\includegraphics[width=0.50\textwidth]{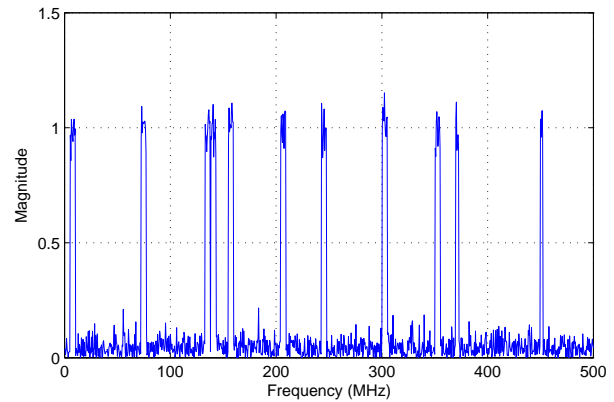}
		\caption{One realization of the noisy spectrum utilization. SNR = 15 dB.}
	\label{fig:alpha_SNR15}
\end{figure}

\begin{figure}
	\centering
		\includegraphics[width=0.50\textwidth]{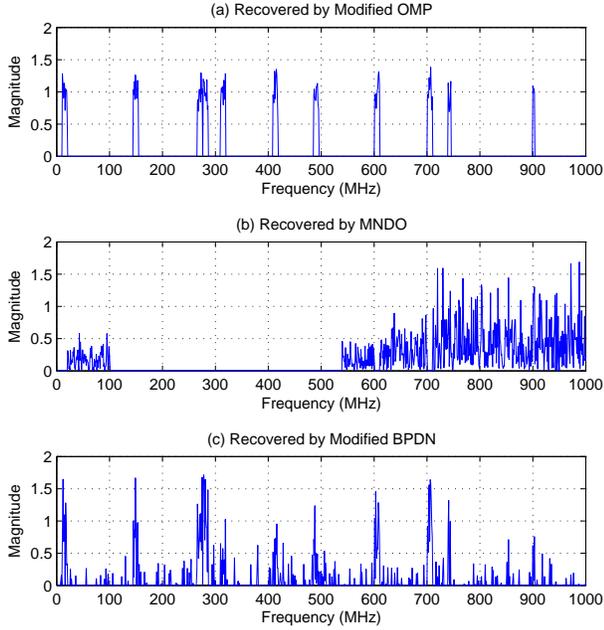}
	\caption{(a) Recovered spectrum by modified OMP. (b) Recovered spectrum by MNDO. (c) Recovered spectrum by modified BPDN.}
	\label{fig:NoisyCase}
\end{figure}

Next we show the example of the algorithms' reconstruction ability under AWGN. The original noisy spectrum is shown in Fig. \ref{fig:alpha_SNR15}, with signal-to-noise ratio (SNR) set to $15$ dB. With $M = 250$, the reconstructed spectrum using these algorithms are shown in Fig. \ref{fig:NoisyCase}. Obviously, modified OMP has best performance among all of them, with MSE $\approx 0.0386$. The non-zero positions are perfectly found and errors are made from the fluctuation of the non-zero frequency values. Other algorithms have similar noisy signal reconstructed as the noiseless example. MNDO has MSE $\approx 2.0007$ and modified BPDN has MSE $\approx 0.4248$.

\begin{figure}
	\centering
		\includegraphics[width=0.50\textwidth]{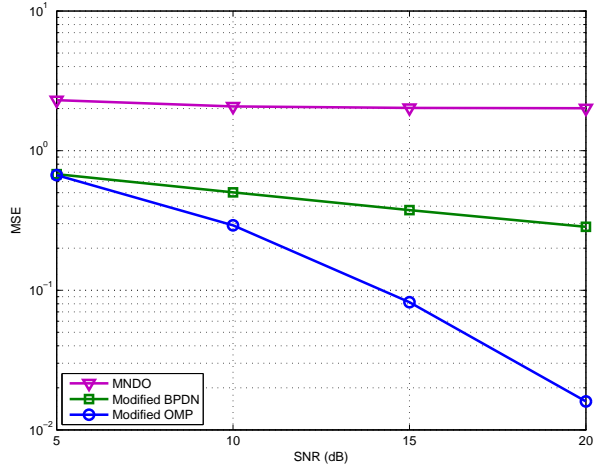}
	\caption{$M = 250$, MSE performances for all algorithms under AWGN.}
	\label{fig:250pts_SNR_MSE}
\end{figure}

\begin{figure}
	\centering
		\includegraphics[width=0.50\textwidth]{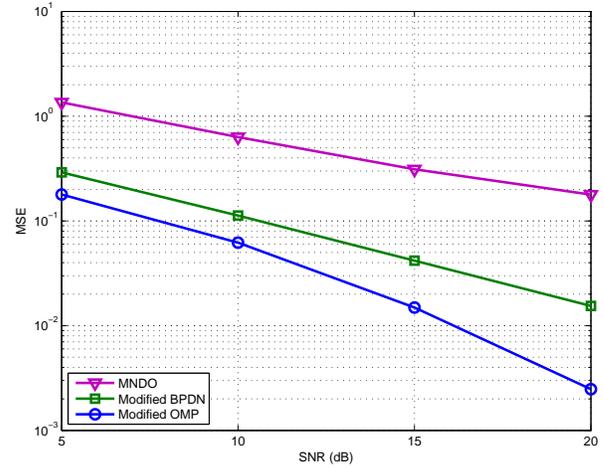}
	\caption{$M = 500$, MSE performances for all algorithms under AWGN.}
	\label{fig:500pts_SNR_MSE}
\end{figure}

Monte Carlo simulation is performed in the noisy signal reconstruction. We use $M = 250$ and $M = 500$ for simulations. SNR is changed from $5$ dB to $20$ dB with $5$ dB increments. Fig. \ref{fig:250pts_SNR_MSE} and Fig. \ref{fig:500pts_SNR_MSE} show the simulation results. For each SNR plot, $500$ noisy signal realizations are used. Obviously, modified OMP outperforms other two BP based algorithms in AWGN as well. Better MSE performance for all algorithms can be achieved by increasing SNR or $M$.

The final test of the reconstruction algorithms is the computation time. All algorithms are written in Matlab and run on a computer with $2.8$ GHz Pentium 4 CPU. The code for modified OMP is written in basic Matlab functions. The codes for MNDO in (\ref{blockmodel}) and modified BPDN in (\ref{modcs}) are written based on \texttt{cvx}, a package for specifying and solving convex programs \cite{cvx,gb08}. To get the results in Fig. \ref{fig:NoiselessCase}, modified OMP takes about 0.05 seconds, MNDO takes about 12.83 seconds, and modified BPDN takes about 5.89 seconds. Obviously, modified OMP is much faster.

\section{Conclusion}
\label{conclusion}
By grouping cognitive radio spectrum usages into three categories, we propose a modified OMP for cognitive radio wideband spectrum sensing. For comparison purpose, we introduce modified BPDN for wideband spectrum sensing as well. The modified OMP is compared with MNDO and modified BPDN, which are BP based. We show by simulation and conclude that, by grouping cognitive radio spectrum usages into three categories, the proposed modified OMP outperforms other two BP based algorithms in both accuracy and computation efficiency. The improvements in accuracy and speed lie in the proper use of prior knowledge and the greedy aspects of the OMP.

Based on the OMP framework, more work is encouraged to be done to explore new algorithms in cooperative spectrum sensing, using reasonable prior knowledge to further improve the performance of CS based wideband spectrum sensing.

\section*{Acknowledgment}
This work is funded by National Science Foundation through grants (ECCS-0901420), (ECCS-0821658) and Office of Naval Research through two contracts (N00014-07-1-0529, N00014-11-1-0006).

\bibliographystyle{ieeetr}
\bibliography{bib/CR,bib/CR_CS,bib/Qiu_Group_bib,bib/Compressed_sensing/Applications/Analog_Information_Conversion,bib/Compressed_sensing/Applications/Astronomy,bib/Compressed_sensing/Applications/Biosensing,bib/Compressed_sensing/Applications/Communications,bib/Compressed_sensing/Applications/Compressive_Imaging,bib/Compressed_sensing/Applications/Compressive_Radar,bib/Compressed_sensing/Applications/Geophysical_Data_Analysis,bib/Compressed_sensing/Applications/Hyperspectral_Imaging,bib/Compressed_sensing/Applications/Medical_Imaging,bib/Compressed_sensing/Applications/Spectrum_analysis,bib/Compressed_sensing/Applications/Surface_metrology,bib/Compressed_sensing/Compressive_Sensing/compressive_sensing,bib/Compressed_sensing/Recovery_Algorithms/recovery_algorithms,bib/Compressed_sensing/Data_Stream_Algorithms/Dimension_Reduction_Embeddings,bib/Compressed_sensing/Data_Stream_Algorithms/Heavy_Hitters,bib/Compressed_sensing/Data_Stream_Algorithms/Histogram_Maintenance,bib/Compressed_sensing/Data_Stream_Algorithms/Random_Sampling,bib/Compressed_sensing/Extensions/extensions,bib/Compressed_sensing/Extensions/multi_sensor_distributed_CS,bib/Compressed_sensing/Foundations_Connections/Bayesian_Methods,bib/Compressed_sensing/Foundations_Connections/Coding_Info_Thry,bib/Compressed_sensing/Foundations_Connections/Finite_Rate_Innovation,bib/Compressed_sensing/Foundations_Connections/High_Dim_Geometry,bib/Compressed_sensing/Foundations_Connections/l1_Norm_Minimization,bib/Compressed_sensing/Foundations_Connections/Lossy_Compression,bib/Compressed_sensing/Foundations_Connections/Machine_Learning,bib/Compressed_sensing/Foundations_Connections/Multiband_Signals,bib/Compressed_sensing/Foundations_Connections/Statistical_Signal_Processing,bib/Compressed_sensing/Tutorials/TUTORIALS}

\end{document}